# Analysis of Magnetic Fields in Inertial Alfvén Wave Collisions

D. J. Drake, J. W. R. Schroeder, B. C. Shanken, G. G. Howes, F. Skiff,
C. A. Kletzing, T. A. Carter, and S. Dorfman

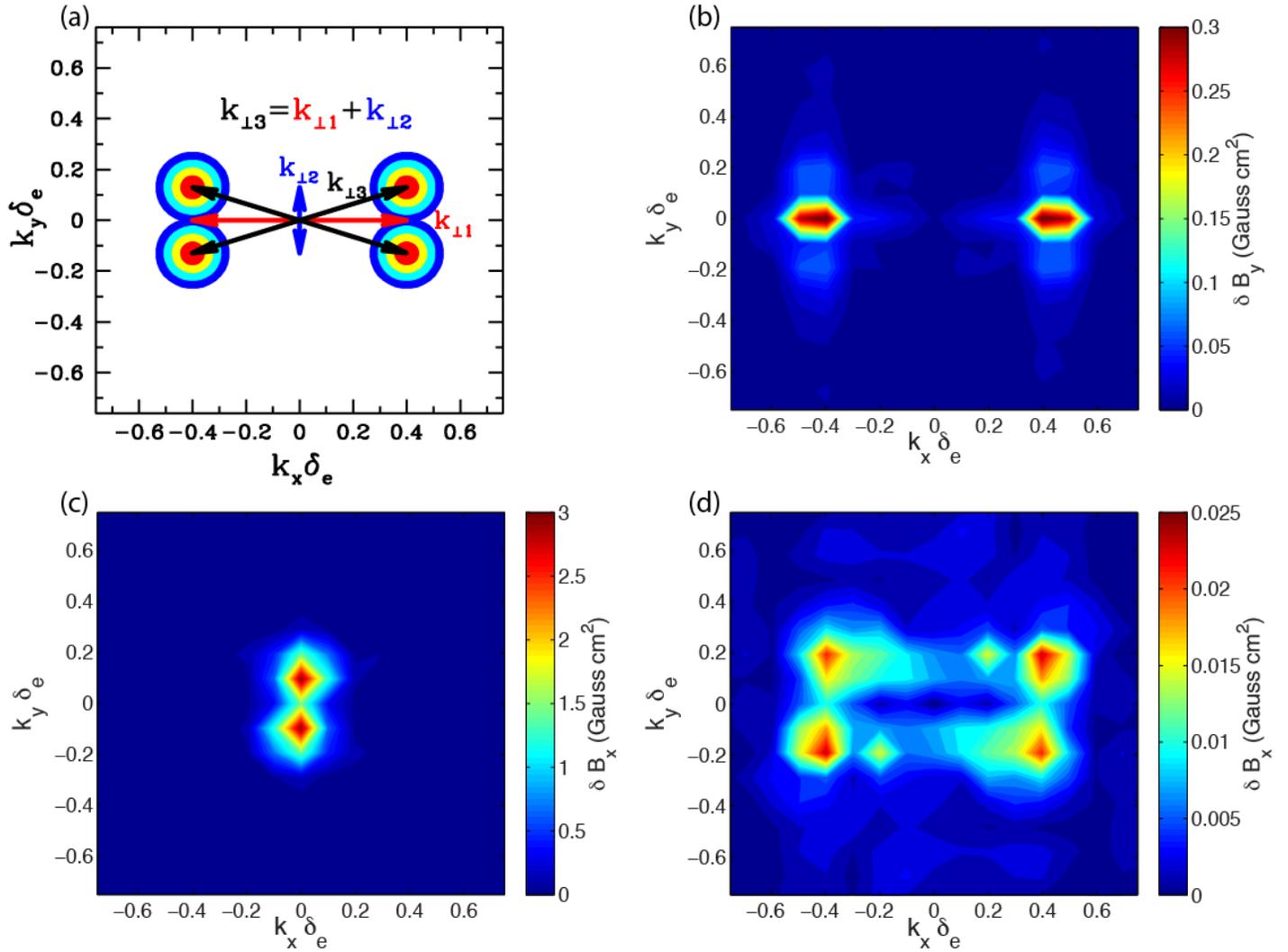

Fig. 1. Colormap plots of the two-dimensional Fourier power spectrum of the (b) $\delta B_y$ component of ASW antenna signal, (c) the $\delta B_x$ component of the loop antenna signal, and (d) the $\delta B_x$ component of the daughter wave signal at $t = 9.66$ ms after the start of the plasma pulse. The daughter wave signal in panel (d) should be a vector sum of two parent waves, $\mathbf{k}_3 = \mathbf{k}_1 + \mathbf{k}_2$. The bulls-eyes in panel (a) shows the predicted value of the daughter wave signal given the perpendicular wave vectors of the two parent waves: $k_{\perp 1}$ (the red line) represents the perpendicular wave vector of the ASW antenna signal and $k_{\perp 2}$ (the blue line) represents the wave vector for the loop antenna signal.

Manuscript received…..
D. J. Drake and B. C. Shanken with Valdosta State University, Department of Physics, Astronomy, and Geosciences, Valdosta, GA 31698.
J. W. R. Schroder, G. G. Howes, F. Skiff, and C. A. Kletzing with University of Iowa, Department of Physics and Astronomy, Iowa City, IA 55242.
T. A. Carter and S. Dorfman with University of California, Department of Physics and Astronomy, Los Angeles, CA 90095-1547.
Work supported by NSF PHY-10033446, NSF CAREER AGS-1054061, NSF CAREER PHY-0547572, and NASA NNX10AC91G.
Publisher Identifier S XXXX-XXXXXXX-X

*Abstract* – Turbulence in astrophysical and space plasmas is dominated by the nonlinear interaction of counterpropagating Alfvén waves. Most Alfvén wave turbulence theories have been based on ideal plasma models for Alfvén waves at large scales. However, in the inertial Alfvén wave regime, relevant to magnetospheric plasmas, how the turbulent nonlinear interactions are modified by the dispersive nature of the waves remains to be explored. Here we present the first laboratory evidence of the nonlinear interaction in the inertial regime. A comparison is made with the theory for MHD Alfvén waves.

Astrophysical plasma turbulence is governed by the nonlinear interaction of counterpropagating Alfvén waves, which have been shown to cause energy cascades from large to small length scales [1]. It is believed that these effects play a key role in the regulation of star formation, heating of the interstellar medium, transport of heat in galaxy clusters, acceleration of the solar wind, and heating of the solar corona. Turbulent fluctuations in the interstellar medium and solar wind have been observed for several decades. Although these measurements provide a good understanding of the effects of turbulence on the plasma environment, they do not provide insight into the physical mechanisms comprising the turbulence [2]. Laboratory experiments, by their very nature, provide a controlled environment in which the small scale turbulent interactions between counterpropagating Alfvén waves can be observed, measured, and understood.

For MHD Alfvén waves, incompressible MHD predicts that a nonlinear interaction between counterpropagating Alfvén waves will occur. This can be seen directly from the equations of incompressible MHD written in the symmetric Elsässer form

$$\frac{\partial \mathbf{z}^\pm}{\partial t} \mp \mathbf{v_A} \cdot \nabla \mathbf{z}^\pm = -\nabla P/\rho_0 - \mathbf{z}^\mp \cdot \nabla \mathbf{z}^\pm. \quad (1)$$

Here the Elsässer fields are given by $\mathbf{z}^\pm = \mathbf{u} \pm \delta \mathbf{B}/\sqrt{4\pi \rho_0}$, where $\mathbf{z}^+$ represents an Alfvén wave traveling parallel to the background magnetic field $\mathbf{B_0}$, and $\mathbf{z}^-$ a wave traveling antiparallel. The Alfvén velocity due to $\mathbf{B_0}$ is given by $\mathbf{v_A}$, $P$ is the total pressure, and $\rho_0$ is the mass density. The last term is the nonlinear interaction term between counterpropagating Alfvén waves. In order for this nonlinear term to be nonzero two conditions must be met: (1) there must be two counterpropagating waves, i.e. $\mathbf{z}^+ \neq 0$ and $\mathbf{z}^- \neq 0$, and (2) they must not be polarized in the same direction, $\mathbf{z}^+ \times \mathbf{z}^- \neq 0$. The lowest order nonlinear interaction between two plane waves, $\mathbf{k_1}$ and $\mathbf{k_2}$, will then be the three-wave interaction that generates a daughter wave, $\mathbf{k_3}$, which satisfies the vector sum $\mathbf{k_1} + \mathbf{k_2} = \mathbf{k_3}$ and $\omega_1 + \omega_2 = \omega_3$. Although the theory [1] has been developed for MHD Alfvén waves based on Eq. (1), an analogous theory for inertial Alfvén waves, $v_A > v_{the}$, should exist with the expectation that the physics of these waves will be different [3].

The experiment [2] reported here was conducted in the Large Plasma Device at UCLA. Two Alfvén wave antennas were positioned at opposite ends of the plasma chamber, ~11 m apart. The UCLA loop antenna produced a sinusoidal waveform with frequency 60 kHz and the Iowa ASW antenna produced a sinusoidal wave of frequency 270 kHz. From a swept Langmuir probe, in conjunction with a microwave interferometer, the electron density in the interaction region was measured as $n_e = 1.0 \times 10^{12}$ cm$^{-3}$ and the electron temperature was $T_e = 4$ eV. With $\mathbf{B_0} = (1800$ G$)\hat{z}$, the Alfvén speed and the electron thermal speed were determined to be $v_A = 3.92 \times 10^8$ cm/s and $v_{the} = 8.38 \times 10^7$ cm/s. This indicates that the Alfvén waves are in the inertial regime. The perpendicular components of the magnetic field were then measured using two Elsässer probes [4]. The measurements were taken over a 30 cm square region, centered on the machine axis.

The spatial Fourier transform of the two antenna signals and the daughter wave signal are shown in Fig. 1. The colormaps in the figure include a linear interpolation to fill the location in the plot between the actual measurements. In Fig. 1(b) we show the 2-D transform of the ASW antenna signal, which shows that the $\delta B_y$ component has perpendicular wave vector of $k_x \delta_e = \pm 0.4$. Fig. 1(c) shows the transform of the $\delta B_x$ component for the loop antenna, which indicates a $k_y \delta_e = \pm 0.1$. The resulting daughter wave is shown in panel (d). Here we can clearly see that the daughter wave signal is the result of a vector sum of the ASW and loop antenna signals, as predicted from Eq. (2) and shown in panel (a). The result is the first experimental evidence of the nonlinear daughter wave produced by the interaction of two counterpropagating inertial Alfvén waves.